\documentclass[twocolumn,pra,aps,showpacs]{revtex4}
\usepackage[dvips]{graphicx}
\usepackage{graphicx}
\usepackage{dcolumn}
\usepackage{bm}

\begin{document}

\title{Tunable lateral shift and polarization beam splitting of the transmitted light beam through electro-optic crystals}

\author{Xi Chen$^{1}$\footnote{Email address: xchen@shu.edu.cn}}

\author{Ming Shen$^{1}$}

\author{Zhen-Fu Zhang$^{1}$}

\author{Chun-Fang Li$^{1,2}$\footnote{Email address: cfli@shu.edu.cn}}

\affiliation{$^{1}$ Department of Physics, Shanghai University,
Shanghai 200444, People's Republic of China}

\affiliation{$^{2}$ State Key Laboratory of Transient Optics and
Photonics, Xi'an Institute of Optics and Precision Mechanics of CAS,
Xi'an 710119, People's Republic of China}


\begin{abstract}
We have investigated the tunable lateral shift and polarization beam
splitting of the transmitted light beam through electro-optic
crystals, based on the Pockels effect. The positive and negative
lateral shifts could be easily controlled by adjusting the
permittivity tensor, which is modulated by the external applied
electric field. An alternative way to realize the polarization beam
splitter was also proposed by the polarization-dependent lateral
shifts. Numerical simulations for Gaussian-shaped incident beam have
demonstrated the above theoretical results obtained by stationary
phase method. All these phenomena have potential applications in
optical devices.



\end{abstract}

\maketitle
\section{INTRODUCTION}

It is well known that a light beam totally reflected from an
interface between two dielectric media undergoes lateral shift from
the position predicted by geometrical optics \cite{Goos}. This
phenomenon was referred to as the Goos-H\"{a}nchen (GH) effect
\cite{Lotsch} and was theoretically explained firstly by Artmann in
1948 \cite{Artmann}. Up till now, the investigations of the GH
shifts have been extended to frustrated total internal reflection
\cite{ Ghatak,Haibel}, and attenuated total reflection
\cite{Yin,Pillon}, and other areas of physics \cite{Lotsch}, such as
quantum mechanics \cite{Renard}, acoustics \cite{Briers}, neutron
physics \cite{Ignatovich}, spintronics \cite{Chen} and atom optics
\cite{Zhang-WP}.

In early 1970s, Reesor \textit{et al.} \cite{Read-1,Read-2} found
that the lateral shift of a light beam incident on an ordinary
dielectric slab is different from the prediction of geometrical
optics, when the slab's thickness is comparable with the wave-length
of light. Hsue and Tamir \cite{Hsue-T} further discussed the lateral
shift of a light beam in a transmitting-layer configuration. But
they concluded that it is always shifted in a forward direction.
Recently, we have predicted theoretically \cite{Li} and demonstrated
experimentally \cite{Li-OC} that the lateral shift can be negative
as well as positive. Historically, the phenomenon of the GH shifts
are usually believed to be associated with the evanescent waves in
total reflection \cite{Goos} and frustrated total internal
reflection \cite{ Ghatak,Haibel}. However, the lateral shift
discussed here has nothing to do with the evanescent wave. The
negative and positive lateral shifts are due to the finite width of
the light beam and are different from the prediction of geometric
optics. So the lateral shift is similar to but different from GH
shift in total reflection, and does result from the reshaping effect
of transmitted beam.

Most recently, large (positive and negative) lateral shifts in
different slabs containing various materials (such as weakly
absorbing media \cite{Lai,Wang-Chen-Zhu}, gain media
\cite{Fan-Wang,Yan}, negative-phase-velocity (NPV) media
\cite{Berman,Lakhtakia-1,Chen-PRE,Wang-Zhu} and anisotropic
metamaterial media \cite{Yiang,Wang-Wang-Zhang}) have attracted much
attention, because of their potential applications in integrated
optics \cite{Lotsch}, electromagnetic communication system
\cite{Li-Vernon}, and optical sensors \cite{Yin-Hesselink,Yu}.
However, it is important for the applications in optical devices to
realize the tunability of lateral shift, that is, to control the
lateral shift in a fixed configuration or device by external field.
Wang \textit{et. al.} \cite{Wang-I-Z} once proposed that the lateral
shift can be modulated by a coherent control field, which is applied
onto the two-level atoms inside a cavity. The modulation of the
lateral shift is significant for the further applications in
flexible optical-beam steering and optical devices in information
processing. Therefore, the main purpose of this paper is to
investigate the control of the lateral shift in electro-optic (EO)
crystals, based on the Pockels effect \cite{Goldstein}, which offers
opportunity for tuning the optical response characteristics of
materials in photonic band-gap engineering \cite{Li-ML,Jim},
composite materials \cite{Lakhtakia,Mackay} and surface-wave
propagation \cite{Nelatury}, due to the linear change of the
refractive indices caused by application of an external electric
field.

Our paper is organized as follows. In Sec. II, we derive the lateral
shifts of TE and TM polarized light beams in transmission through
the EO crystals, according to the stationary phase approach. In Sec.
III, we discuss the electric control of the lateral shifts in the
case of different crystal cuts. In the same section, an alternative
way to realize the polarization beam splitter is also proposed. In
Sec. IV, numerical simulations for Gaussian-shaped beam are made to
demonstrate the validity of above theoretical results given by
stationary phase method. Finally, a conclusion will be given in Sec.
V.

\section{FORMULA}
\label{Theory}

\begin{figure}[]\begin{center}
\scalebox{0.8}[0.8]{\includegraphics{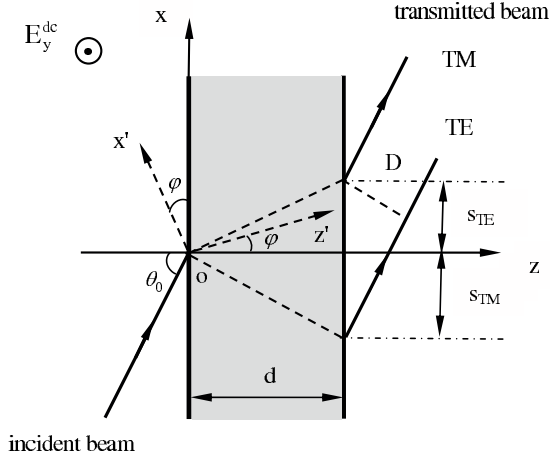}} \caption{Schematic
diagram of positive and negative lateral shifts of TE and TM
polarized light beams in transmission through a slab of EO crystal
with external applied electric field in the $y$ direction.}
\label{configuration}
\end{center}\end{figure}

Consider a light beam of angular frequency $\omega$ incident on a
slab of EO crystal in the air with an incidence angle $\theta_0$
specified by the inclination of the beam with respect to the $z$
axis, as shown in Fig. \ref{configuration}, where the thickness,
relative permittivity and relative permeability
of the nonmagnetic EO crystal slab, are denoted by $d$,
$\hat{\varepsilon}_1$, and $\mu_0$, respectively. In the case of TE
(TM) polarization, the electric (magnetic) field of the plane wave
component of the incident beam is assumed to $
\Psi_{in}(\vec{x})=A\exp(i \vec{k}\cdot \vec{x})$, where $\vec{k}
\equiv (k_x, k_z)=(k\sin\theta, k\cos\theta)$, $k= (\varepsilon_0
\mu_0)^{1/2}\omega/c$ is the wave number in the air,
$\varepsilon_0$, and $\mu_0$ are the relative permittivity and
permeability of the air, $c$ is the speed of light in vacuum, and
$\theta$ stands for the incidence angle of the plane wave under
consideration. For the sake of simplicity, suppose we have a
tetragonal (point group $\bar{4}2 \mbox{m}$) uniaxial crystal. The
EO crystal with optic axis in the $z$ direction has the following
index ellipsoid equation:
\begin{equation}
\frac{x^2}{n_o^2}+\frac{y^2}{n_o^2}+\frac{z^2}{n_e^2}=1,
\end{equation}
and the relative permittivity tensor is,
\begin{eqnarray}
\hat{\varepsilon}_1=\left(\begin{array}{ccc}\varepsilon_x & 0 & 0 \\ 0 & \varepsilon_x & 0 \\ 0 & 0 & \varepsilon_z \\
\end{array}
\right),
\end{eqnarray}
where $\varepsilon_x=n^2_{o}$, $\varepsilon_z=n^2_{e}$, $n_{o}$ and
$n_{e}$ are the refractive indices for ordinary and extraordinary
waves inside the anisotropic slab, respectively. In a electric field
$E^{dc}_y$ along the direction of $y$, the index ellipsoid equation
becomes
\begin{equation}
\label{index equation}
\frac{x^2}{n_o^2}+\frac{y^2}{n_o^2}+\frac{z^2}{n_e^2}+2\gamma_{41}E^{dc}_y
xz=1.
\end{equation}
By making coordinate transformations, as shown in Fig.
\ref{configuration},
\begin{equation}
\label{transformations}
\left(\begin{array}{c}x' \\ z' \\
\end{array}\right)=\left(\begin{array}{ccc} \cos\varphi & -\sin\varphi \\ \sin \varphi & \cos\varphi  \\
\end{array}\right)\left( \begin{array}{c}x \\ z \\
\end{array}\right)
\end{equation}
the index ellipsoid equation (\ref{index equation}) is obtained by
\begin{equation}
\frac{x'^2}{n^{2}_{x'}} +\frac{y'^2}{n^{2}_{y'}}
+\frac{z'^2}{n^{2}_{z'}}=1,
\end{equation}
where
\begin{eqnarray*}
n^{2}_{x'}\equiv\varepsilon'_x=\left(\frac{1}{n_o^2}\cos^2\varphi+\frac{1}{n_e^2}\sin^2\varphi+\gamma_{41}E^{dc}_y\sin2\varphi\right)^{-1},
\end{eqnarray*}
\begin{eqnarray*} n^{2}_{y'}\equiv\varepsilon'_y=n^{2}_{o},\end{eqnarray*}
and
\begin{eqnarray*}
n^{2}_{z'}\equiv\varepsilon'_z=\left(\frac{1}{n_o^2}\sin^2\varphi+\frac{1}{n_e^2}\cos^2\varphi-\gamma_{41}E^{dc}_y\sin2\varphi\right)^{-1},
\end{eqnarray*}
when the angle $\varphi$ is determined by
\begin{equation}
\tan2\varphi=-\frac{2\gamma_{41}E^{dc}_y}{\frac{1}{n_e^2}-\frac{1}{n_o^2}}.
\end{equation}
Moreover, when the optic axis is changed to be in the $y$ direction,
the index ellipsoid equation and relative permittivity tensor can be
expressed by
\begin{equation}
\frac{x^2}{n_o^2}+\frac{y^2}{n_e^2}+\frac{z^2}{n_o^2}=1,
\end{equation}
and
\begin{eqnarray}
\hat{\varepsilon}_1=\left(\begin{array}{ccc}\varepsilon_x & 0 & 0 \\ 0 & \varepsilon_y & 0 \\ 0 & 0 & \varepsilon_x \\
\end{array}
\right),
\end{eqnarray}
respectively. The index ellipsoid equation in a field along the
direction of $y$ becomes
\begin{equation}
\label{index equation-2}
\frac{x^2}{n_o^2}+\frac{y^2}{n_e^2}+\frac{z^2}{n_o^2}+2\gamma_{63}E^{dc}_y
xz=1.
\end{equation}
Let $\varphi=-\pi/4$ in the coordinate transformations
(\ref{transformations}), the above index ellipsoid equation is given
by
\begin{equation}
\frac{x'^2}{n^{2}_{x'}} +\frac{y'^2}{n^{2}_{y'}}
+\frac{z'^2}{n^{2}_{z'}}=1,
\end{equation}
where $n_{x'}=1/(1/n_o^2-\gamma_{63}E^{dc}_y )$, $n_{y'}=n_{e}$, and
$n_{z'}=1/(1/n_o^2+\gamma_{63}E^{dc}_y )$. As EO coefficients
$\gamma_{41}$ and $\gamma_{63}$ are quite different for the
tetragonal uniaxial crystals such as ADP and KDP with the others
$\gamma_{ij}=0$, it clearly makes sense to utilize the two
coefficients mentioned above to control the lateral shift, taking
account into different crystal cuts. All these expressions obtained
here are also valid for the cubic crystal of point group
$\bar{4}3\mbox{m}$, the symmetry group of such common materials as
GaAs, where $n_o=n_e$, $\gamma_{41}=\gamma_{63}$.

In order to calculate the lateral shifts, the new relative
permittivity tensors in the original $xyz$ frame is expressed as the
following form:
\begin{eqnarray}
\hat{\varepsilon'} = \left(\begin{array}{ccc} a & 0 & f \\
0 & \varepsilon_y' & 0 \\ f & 0 & b \\
\end{array} \right),
\end{eqnarray}
where $a=\varepsilon_x'\cos^2\varphi+\varepsilon_z'\sin^2\varphi$,
$b=\varepsilon_x'\sin^2\varphi+\varepsilon_z'\cos^2\varphi$,
$f=(\varepsilon_z'-\varepsilon_x')\sin\varphi\cos\varphi$. Then, the
dispersion equations are given by
\begin{equation}
\label{dispersion TE-2} k^{2}_x+k{'}^{2}_z= \varepsilon'_y \mu_0
\frac{\omega^2}{c^2}, ~~~ (\mbox{for TE wave}),
\end{equation}
and
\begin{equation}
\label{dispersion TM-2} k{'}^{2}_z + 2 \frac{f}{b} k_x k{'}_z +
\frac{a}{b} k^{2}_x= \varepsilon'_x \varepsilon'_z \frac{\mu_0}{b}
\frac{\omega^2}{c^2}, ~~ (\mbox{for TM wave}).
\end{equation}
where $k{'}_{z}$  for TM polarized wave is also expressed by
\begin{equation}
k{'}_{z\pm}= -\alpha_1 \pm \alpha_2,
\end{equation}
where $\alpha_{1} = f k_{x}/b$, $\alpha_{2} =  \sqrt{\gamma}k /b$,
$\gamma=\varepsilon'_x \varepsilon'_z (b \mu_0-\sin^2 \theta)$.
Thus, the field of the corresponding transmitted plane wave is
found, according to Maxwell's equations and the boundary conditions,
to be $ \Psi_t(\vec{x}) = T A \exp \{i[k_x x+k_z (z-d)]\}$, where
the transmission coefficient, $T=e^{i (\phi-\alpha_1 d)}/f$, is
determined by the following complex number,
\begin{equation}
f e^{i \phi}=\cos \alpha_2 d+ \frac{i}{2}\left(\chi \frac{
k_z}{\alpha_2}+ \frac{1}{\chi} \frac{\alpha_2}{ k_z}\right)\sin
\alpha_2 d,
\end{equation}
and
\begin{equation}
\tan \phi = \frac{1}{2}\left( \chi \frac{ k_z}{\alpha_2}+
\frac{1}{\chi} \frac{\alpha_2}{k_z}\right) \tan \alpha_2 d,
\end{equation}
where $\chi=(\varepsilon'_x \varepsilon'_z)/(b \varepsilon_0)$.
Clearly, the phase shift of the transmitted beam at $z=d$ with
respect to the incident beam at $z=0$ is equal to $\phi-\alpha_1 d$.
For a well-collimated TM polarized light beam, the lateral shift is
defined as $-d (\phi-\alpha_1 d)/dk_x|_{\theta=\theta_{0}}$,
according to the stationary phase approach \cite{Artmann,Li}, and is
finally given by
\begin{equation}
\label{lateral shift-TM} s_{TM}= s+ \widetilde{s},
\end{equation}
where
\begin{widetext}
\begin{eqnarray}
s &=& \frac{d \tan \theta'_{0}}{2f_0^{2}} \left(\frac{\varepsilon'_x
\varepsilon'_z}{b^2}\right)\left[\left(\chi \frac{
k_{z0}}{\alpha_{20}}+\frac{1}{\chi}\frac{\alpha_{20}}{k_{z0}}\right)\right.
- \left.\left(1-\frac{b^2}{\varepsilon'_x
\varepsilon'_z}\frac{\alpha^2_{20}}{k^2_{z0}}\right)\left(\chi
\frac{
k_{z0}}{\alpha_{20}}-\frac{1}{\chi}\frac{\alpha_{20}}{k_{z0}}\right)\frac{\sin2
\alpha_{20} d}{2 \alpha_{20} d}\right],
\end{eqnarray}
\end{widetext}
$\widetilde{s}=(f/b) d $, and $\tan \theta'_{0} =
k_{x0}/\alpha_{20}$. It is noted that the subscript $0$ in this
paper denotes values taken at $k_x=k_{x0}$, namely,
$\theta=\theta_0$. Similarly, the transmission coefficient, $T=e^{i
\phi}/f$, of the TE polarized light beam is determined by the
following complex number,
\begin{equation}
fe^{i \phi}=\cos k'_z d+ \frac{i}{2}\left(\chi \frac{ k_z}{k'_z}+
\frac{1}{\chi} \frac{k'_z}{ k_z}\right)\sin k'_z d,
\end{equation}
where the phase shift is given by
\begin{equation}
\label{phase shift} \tan \phi = \frac{1}{2}\left(\chi \frac{
k_z}{k'_z}+ \frac{1}{\chi} \frac{k'_z}{ k_z} \right)\tan k'_z d,
\end{equation}
and $\chi=1$. The lateral shift of the TE polarized light beam is
defined as $-d \phi/dk_x|_{\theta=\theta_{0}}$, and is given by
\begin{widetext}
\begin{eqnarray}
\label{lateral shift-TE} s_{TE} &=& \frac{d  \tan
\theta'_0}{2f_0^{2}}\left[\left(\chi \frac{
k_{z0}}{k'_{z0}}+\frac{1}{\chi}\frac{k'_{z0}}{k_{z0}}\right)\right.
- \left.\left(1-\frac{k'^{2}_{z0}}{k^2_{z0}}\right)\left(\chi \frac{
k_{z0}}{k'_{z0}}-\frac{1}{\chi}\frac{k'_{z0}}{k_{z0}}\right)\frac{\sin2k'_{z0}
d}{2k'_{z0} d}\right],
\end{eqnarray}
\end{widetext}
where $\tan \theta'_0=k_{x0}/k^{'}_{z0}$. The lateral shifts of TM
and TE polarized light beams presented here depends not only on
$\theta_0$ and $d$,  but on $\varepsilon'_x$, $\varepsilon'_z$, and
$\varphi$. From Eqs. (\ref{lateral shift-TM}) and (\ref{lateral
shift-TE}), it is shown that the lateral shifts can be negative when
incidence angle is larger than the threshold of angle \cite{Li}. In
what as follows, the properties of the lateral shifts will be
discussed in detail. For the tetragonal uniaxial crystals, we have
$n_0=1.5266$, $n_e=1.4808$, and $\gamma_{41}=23.76 \times 10^{-12}
\mbox{m/V}$, $\gamma_{63}=8.56 \times 10^{-12} \mbox{m/V}$ for ADP
and $n_0=1.5115$, $n_e=1.4698$, and $\gamma_{41}=8.77 \times
10^{-12} \mbox{m/V}$, $\gamma_{63}=10.3 \times 10^{-12} \mbox{m/V}$
for KDP at $\lambda=546 \mbox{nm}$ \cite{Goldstein}, respectively.

\begin{figure}[]\begin{center}
\scalebox{0.4}[0.4]{\includegraphics{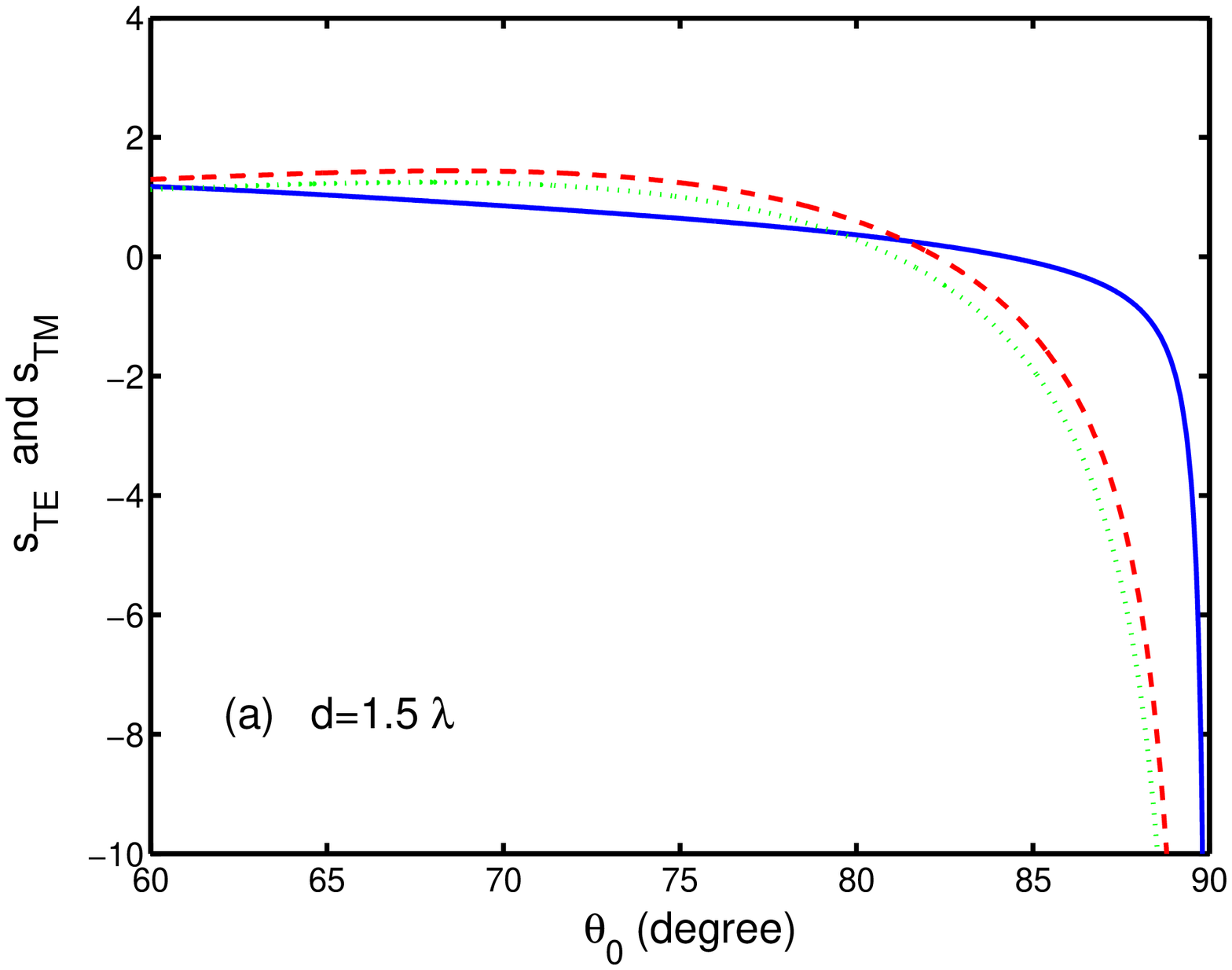}}
\scalebox{0.4}[0.4]{\includegraphics{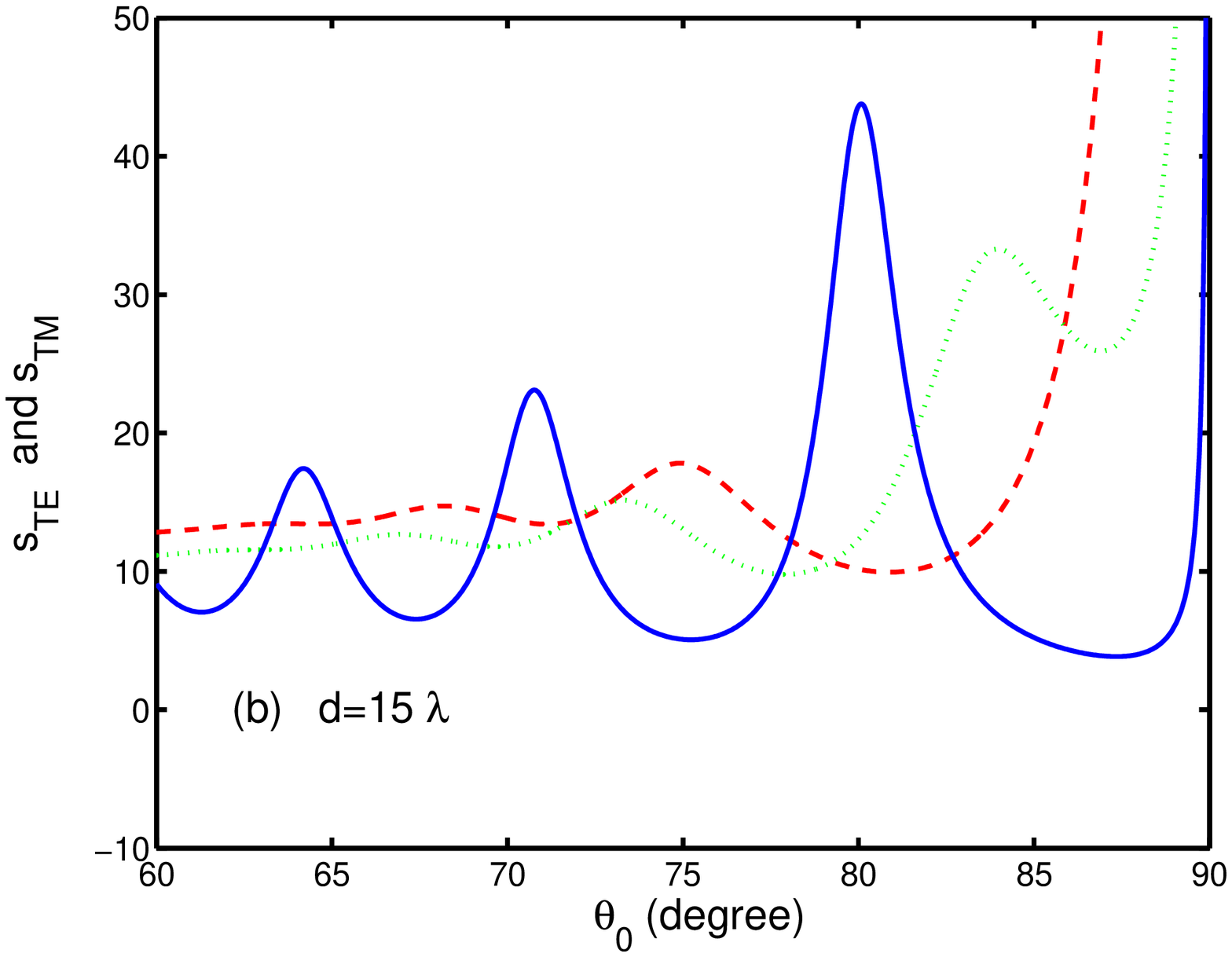}} \caption{Dependence
of the lateral shifts (in units of $\lambda$) on the incidence angle
$\theta_0$ in the slab of ADP crystal with the optic axis in the $z$
direction, where (a) $d=1.5\lambda$ and (b) $d=15 \lambda$. Solid
curve represents $s_{TE}$, dotted and dashed curves represent
$s_{TM}$ corresponding to $E^{dc}_y=0$ and $E^{dc}_y=1 \mbox{GV/m}$,
respectively.} \label{lateral shift}
\end{center}\end{figure}

\section{DISCUSSIONS}

Fig. \ref{lateral shift} shows the dependence of the lateral shifts
of TM and TE polarized light beams on the incidence angle $\theta_0$
in the electric-optic crystal with the optic axis in the $z$
direction, where (a) $d=1.5 \lambda$ and (b) $d=15 \lambda$. Solid
curve represents the lateral shift of TE polarized beam, dotted and
dashed curves represent the lateral shifts of TM polarized beam
under the applied electric field $E^{dc}_y=0$ and $E^{dc}_y=1
\mbox{GV/m}$, respectively. It is shown in Fig. \ref{lateral shift}
that the lateral shift can be negative when the slab's thickness is
small, as mentioned above. The lateral shifts of TM and TE polarized
light beams are quite different in the EO crystals with the optical
axis in the direction $y$ axis. More importantly, the lateral shift
of TM polarized beam can be controlled by the applied electric
field, while the lateral shift of TE polarized beam in this case is
independent of the applied electric field, due to the symmetry of
point group considered here.

\begin{figure}[]\begin{center}
\scalebox{0.4}[0.4]{\includegraphics{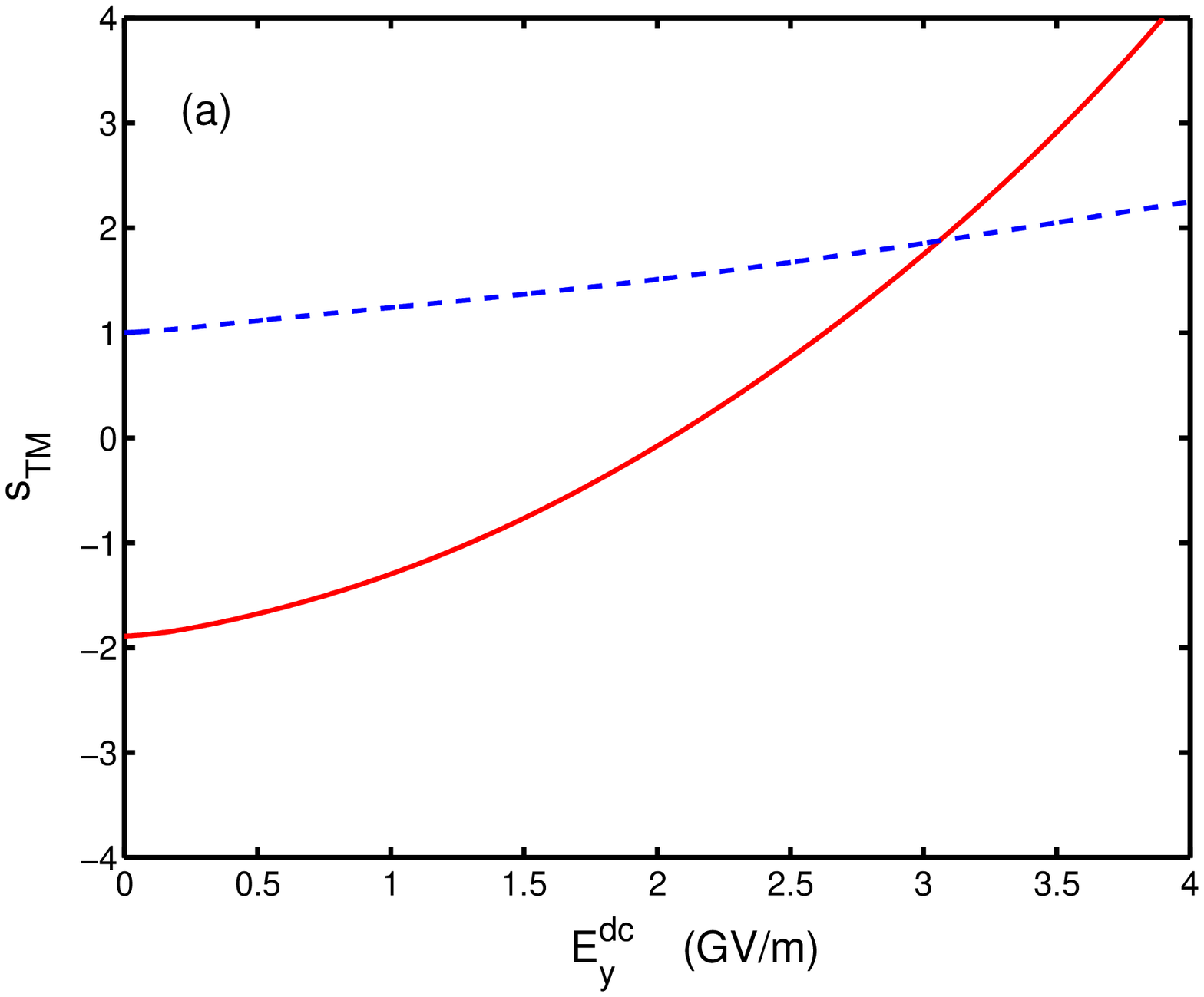}}
\scalebox{0.4}[0.4]{\includegraphics{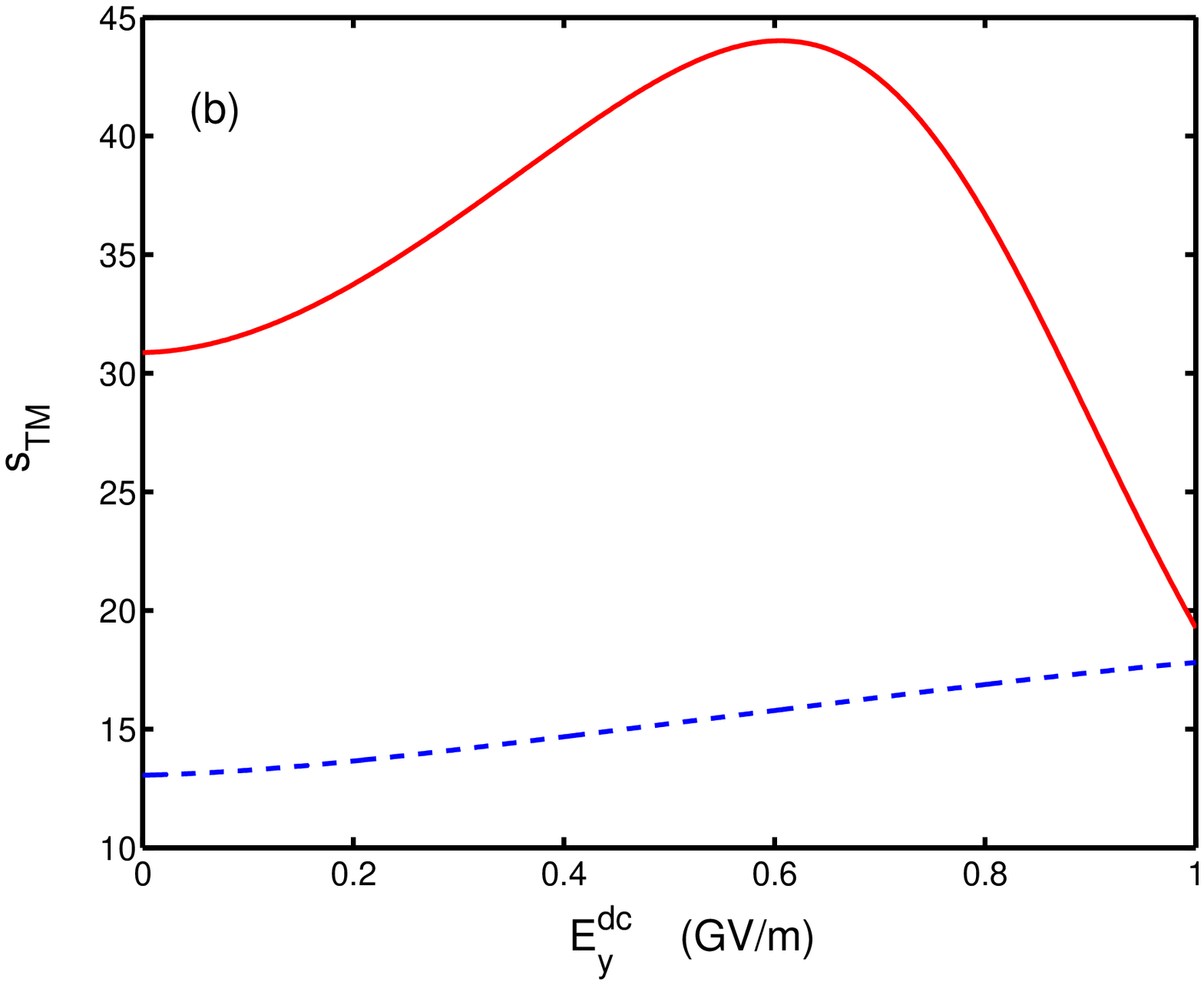}} \caption{Dependence
of the lateral shifts (in units of $\lambda$) on the electric field
at incidence angle $\theta_0 =75^{\circ}$ (dashed curves) and
$\theta_0 =85^{\circ}$ (solid curve), where (a) $d=1.5 \lambda$ and
(b) $d=15 \lambda$, and other parameters are the same as in Fig.
\ref{lateral shift}.} \label{lateral shift-2}
\end{center}\end{figure}

Now, we will discuss the effect of applied electric field on the
lateral shifts. Fig. \ref{lateral shift-2} illustrates that the
lateral shifts of TM polarized beam is dependent on the applied
electric fields at different incidence angles, where all the
physical parameters are the same as in Fig. \ref{lateral shift}. It
is due to the fact that the lateral shift for TM polarized light
beam presented here is closely related to $\varepsilon'_x$,
$\varepsilon'_z$, and $\varphi$, which can be modulated by applied
electric field, based on the Pockles effect. As shown in Fig.
\ref{lateral shift-2}, the lateral shift can be tuned from positive
to negative by the applied electric field at larger incidence angle,
when the slab's thickness is small. It is also shown that the
electric control of the lateral shift can be realized more easily
with the increasing of slab's thickness. Taking account into that
the electrode configuration used for applying an electric field
consists of two coplanar electrodes separated by a $20 \mu m$ wide
gap \cite{Jim}, we can easily control the lateral shifts by applied
electric field when the EO crystal is subjected to a dc voltage
ranging from $0$ to $20 \mbox{KV}$ ($E^{dc}_{y} \sim 1
\mbox{GV/m}$), once one chooses the structure. From a practical
point of view, we should point out that the stronger dc field could
cause an EO device to malfunction seriously, so that we should find
the crystal with large EO coefficients to realize the modulation of
the lateral shift by the finite electric field in magnitude.

\begin{figure}[]\begin{center}
\scalebox{0.4}[0.4]{\includegraphics{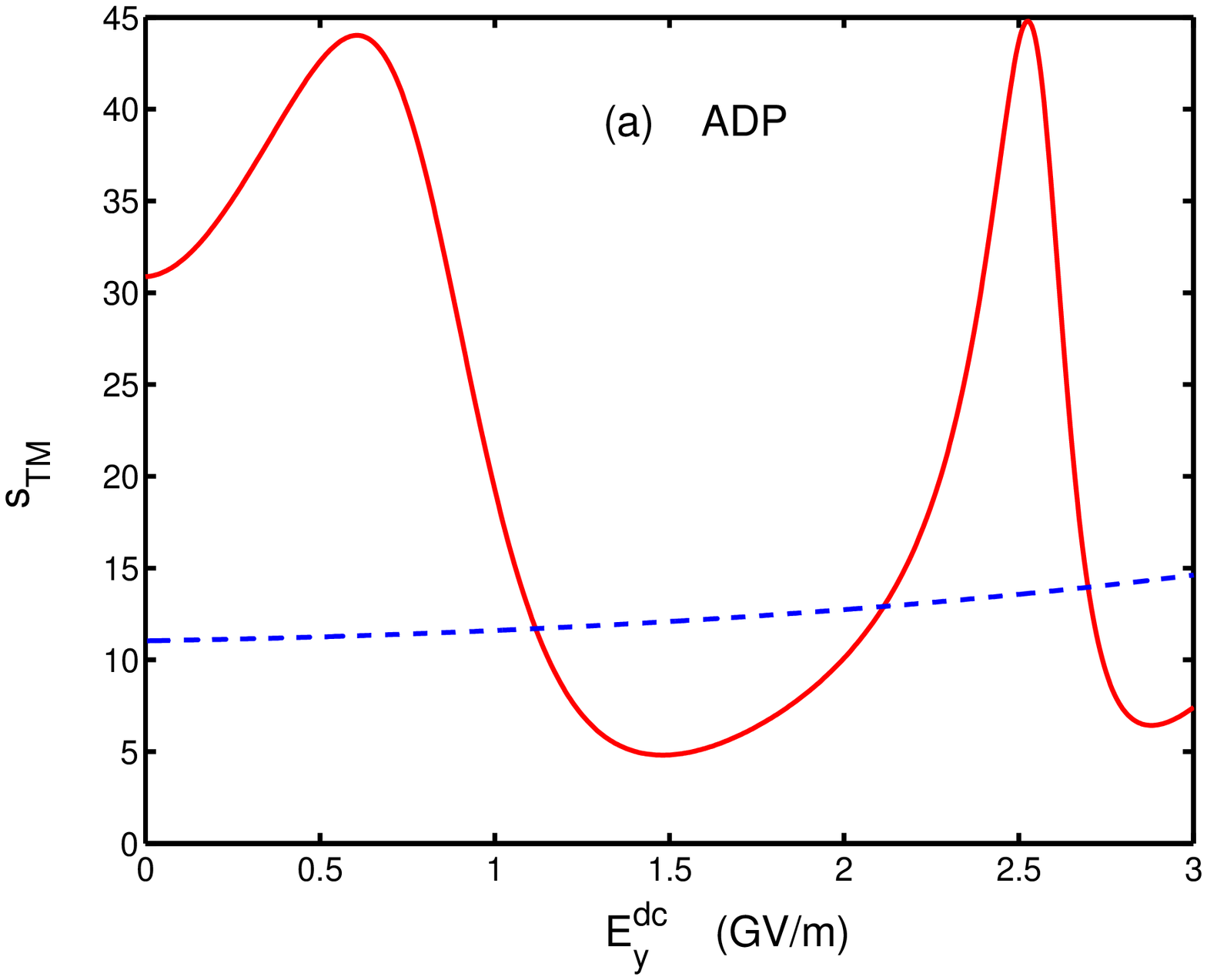}}
\scalebox{0.4}[0.4]{\includegraphics{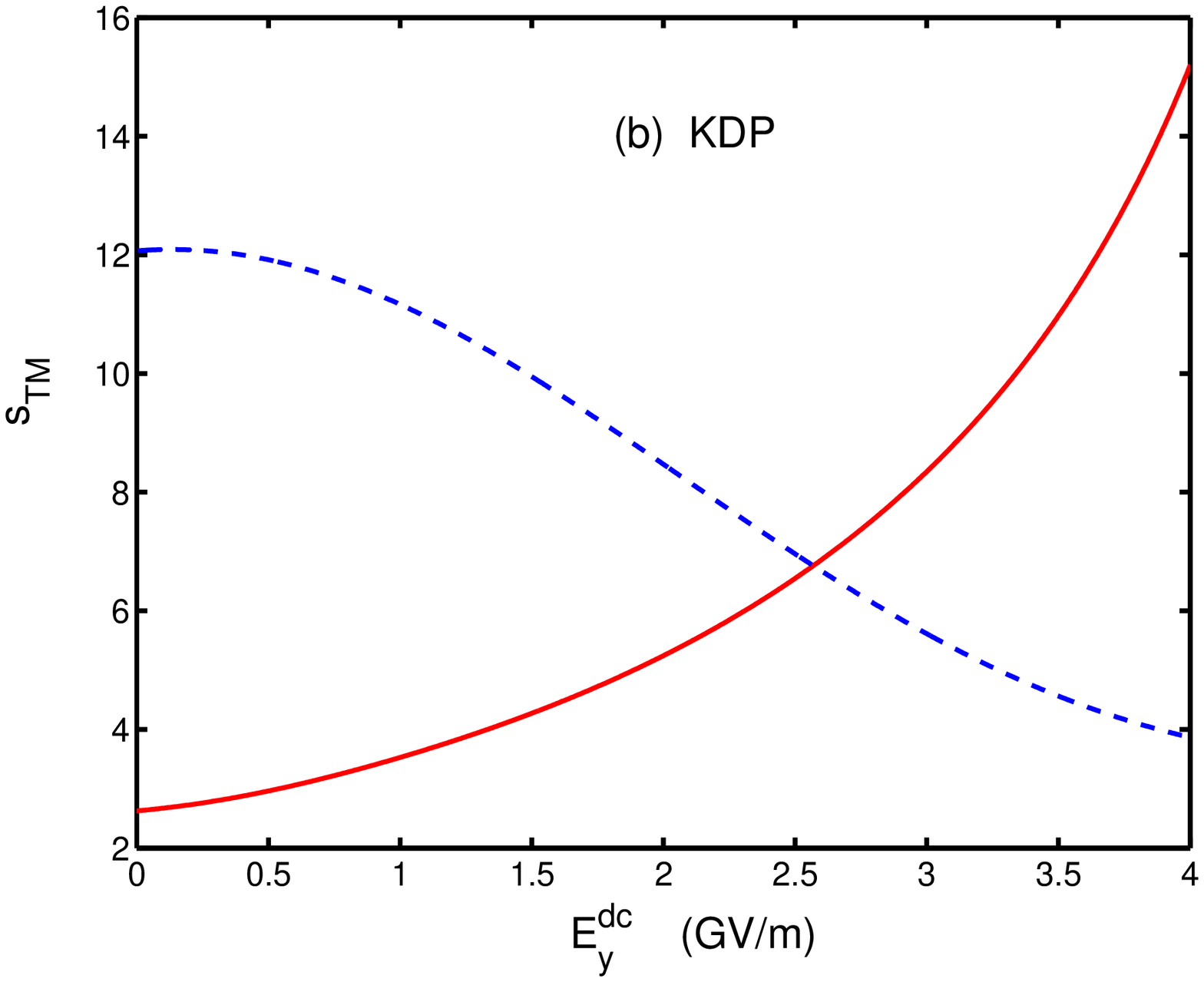}} \caption{Dependence
of the lateral shifts on the electric field in case of different
crystal cuts for (a) ADP  and (b) KDP, where $d=15 \lambda$ and
$\theta_0=85^\circ$. Solid and dashed curves correspond to the
$z$-cut and $y$-cut crystals, respectively.} \label{lateral shift-3}
\end{center}\end{figure}

Fig. \ref{lateral shift-3} further shows theoretically the electric
control of the lateral shifts in the case of different crystal cuts.
Since the modification of the permittivity tensor by the applied
electric field is quantified through $18$ EO coefficients, not all
of which may be independent of each other, depending on the point
group symmetry \cite{Goldstein}, the lateral shifts depend on the
different EO coefficients relating to crystal cuts, as shown in Fig.
\ref{lateral shift-3}, where solid and dashed curves correspond to
the $z$-cut and $y$-cut crystals, respectively. By comparison
between Fig. \ref{lateral shift-3} (a) and (b), it is thus implied
that we can choose the crystal geometry in terms of the relative
large EO coefficients used to control the lateral shift efficiently
when the electrode is fixed.

\begin{figure}[]\begin{center}
\scalebox{0.4}[0.4]{\includegraphics{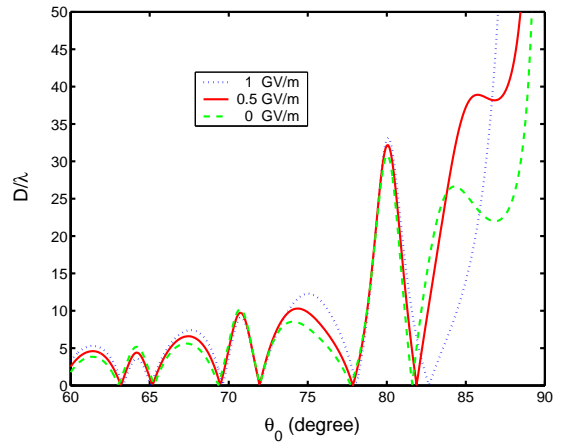}} \caption{Dependence
of $D$ on the incident angle $\theta_0$ under different applied
electric fields, where $d=15 \lambda$, and the other parameters are
the same as in Fig. \ref{lateral shift-2}. \label{D}}
\end{center}\end{figure}

Next, we have a brief look at the polarization beam splitter via
tunable lateral shift. In the above discussions, it is clear that
the lateral shifts depends on the polarization state of the beam,
which leads to the separation of the two orthogonal polarizations of
light beam. To illustrate the influence of the applied electric
field on polarization beam splitter in EO crystal, we define the
distance, as shown in Fig. \ref{configuration},
\begin{eqnarray} D=|s_{TM}-s_{TE}| \times \sin \theta_0, \end{eqnarray}
which depends on the applied electric field, as shown in Fig.
\ref{D}. Clearly, the distance $D$ can be tuned by the applied
electric field. The polarization beam splitting can also be enhanced
by external electric field at some large angles, as shown in Fig.
\ref{D}, where the slab's thickness is chosen to be $d= 10 \lambda$.
Using this phenomena, the high efficient ultra-compact polarization
beam splitter can be achieved by lateral shifts modulated by the
applied electric field.

\section{NUMERICAL SIMULATIONS}

To show the validity of the above stationary-phase analysis,
numerical calculations are performed in this section, which confirm
our theoretical results. In the numerical simulation, an incident
Gaussian-shaped light beam is assumed,
$\Psi_{in}(\vec{x})|_{x=0}=\exp(-x^2/2w^2_x+ik_{x0}x)$, which has
the Fourier integral of the following form,
\begin{equation}
\label{incident beam}
\Psi_{in}(\vec{x})|_{z=0}=\frac{1}{\sqrt{2\pi}}\int_{-\infty}^{+\infty}
A(k_x)\exp(i k_x x)dk_x,
\end{equation}
where $w_x=w \sec\theta_0$, $w$ is the local waist of beam at $z=0$,
and the amplitude angular-spectrum distribution is Gaussian,
$A(k_x)=w_x \exp[-(w^2_x/2)(k_x-k_{x0})^2]$. Consequently, when the
angular-spectrum distribution $A(k_x)$ is sharply around $k_{x0}$,
the  field of the transmitted beam can be written as
\begin{equation}
\label{transmitted beam} \Psi_{t}(\vec{x})=\frac{1}{\sqrt{2 \pi}}
\int_{-\infty}^{+\infty} T A(k_x)\exp\{i[k_x x+k_z(z-d)]\}dk_x.
\end{equation}
If the incident light beam is well collimated, $A(k_x)$ is a sharply
distributed Gaussian function around $k_{x0}$. In this case, the
transmission coefficient $T$ can be approximated, by writing it in
an exponential form, expanding the exponent in Taylor series at
$k_{x0}$ and retaining the first two terms, to be
\begin{equation}
\label{exponet expand} T(k_x) \approx T_0
\exp{\left[\frac{1}{T_0}\frac{dT}{dk_{x0}}(k_x-k_{x0})\right]}.
\end{equation}
Substituting Eq. (\ref{exponet expand}) into Eq. (\ref{transmitted
beam}) and and using paraxial approximation, $k_z \approx k_{z0} - (
k_x -k_{x0}) \tan \theta_0$, we finally get the transmitted beam
$z=d$ as follow,
\begin{equation}
\label{field of transmitted beam} \Psi_t(x, d) \simeq T(k_{x0})
\exp{\left[-\frac{(x-s)^2}{2 w^2_x}\right]} \exp{\left[i
(k_{z0}+\frac{\eta}{w^2_x})z\right]} ,
\end{equation}
where $s$ and $\eta$ are determined by,
\begin{equation} s + i \eta = \frac{i}{T(k_{x0})}\frac{dT}{dk_{x0}}.\end{equation}
Obviously, the lateral shift $s$ is exactly equal to $-\partial
\Phi/\partial k_{x0}$ obtained by stationary phase approach. It is
also found that $\eta/w^2_x$ means that the angle deflection of
propagation direction depends closely on the width of beam waist,
which will disappear when the beam waist is very wide
\cite{Li,Li-OC}.

\begin{figure}[]
\scalebox{0.4}[0.4]{\includegraphics{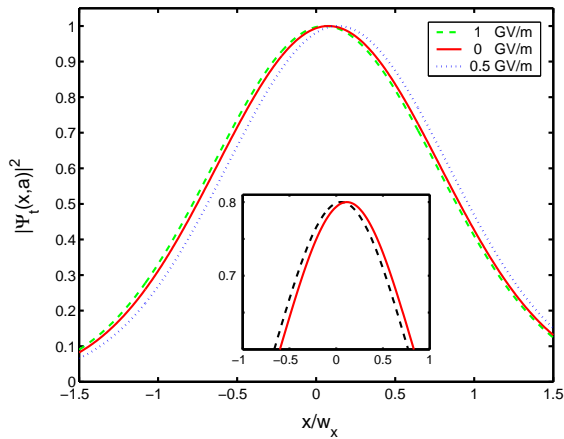}}
\caption{\label{fig.6} Normalized Gaussian shapes of the transmitted
beam through the EO crystal, where $w=150\lambda$, $\theta_0=
85^\circ$, $d = 15 \lambda$, and the other parameters are the same
as Fig. \ref{lateral shift-2}. Dashed curve corresponds to the TM
polarized beam, and solid curves correspond to the TM polarized
beams under the electric field $E^{dc}_y=0$, $E^{dc}_y= 0.5
\mbox{GV/m}$, and $E^{dc}_y= 1 \mbox{GV/m}$, respectively. Inset
shows the polarization beam splitting in the case of $E^{dc}_y=0$.}
\end{figure}

Fig. \ref{fig.6} shows that the normalized Gaussian shapes of the
transmitted beam through the EO crystal, where $w=150\lambda$,
$\theta_0= 85^\circ$, $d = 15 \lambda$, and the other parameters are
the same as Fig. \ref{lateral shift-2}. It is shown that the lateral
shift of TM polarized light beam can be controlled by different
applied electric fields. The comparison of the normalized Gaussian
shapes of the transmitted TE and TM polarized beams has also shown
the polarization beam splitting in the case of $E^{dc}_y=0$, as
shown by inset in Fig. \ref{fig.6}. Therefore, numerical simulations
for a Gaussian-shaped incident beam have demonstrated the validity
of the stationary phase method. The numerical results are in good
agreement with the above theoretical results when the incident beam
is well-collimated, that is, the beam waist of incident beam is
wide. The discrepancy between theoretical and numerical results is
due to the distortion of the transmitted light beam, especially when
the beam waist is narrow.

\section{CONCLUSIONS}

In summary, we have investigated the tunable lateral shifts and
polarization beam splitter based on the Pockels effect in cubic and
tetragonal EO crystals. It is found that the large positive and
negative lateral shifts of TM polarized beam can be controlled by
adjusting the permittivity tensor, which is modulated by external
electric field. It is also shown that the modulation of the lateral
shifts is also closely related to the crystal cuts. Numerical
simulations for Gaussian-shaped light beams are made to demonstrate
the validity of stationary phase approach. Since the EO coefficient
of crystal depends on its symmetry of point group, the lateral
shifts of TE and TM polarized beams can be easily controlled in
orthorhombic crystals of point group $2 \mbox{mm}$, such as
potassium niobate with large EO coefficients $\gamma_{42}=450 \times
10^{-12} \mbox{pm/V}$ and $\gamma_{51}=120 \times 10^{-12}
\mbox{m/V}$ \cite{Lakhtakia}. In a word, we propose theoretically a
useful scheme to control the lateral shifts in EO crystal via the
applied electric field. We believe that these phenomena will lead to
an alternative way to realize polarization beam splitter, which is
essential optical components in optical systems and plays an
important role in optical communication, optical recording and
integrated optical circuits.

\section*{Acknowledgements}
X. Chen is grateful to Prof. A. Lakhtakia for providing useful
information and helpful suggestions. This work was supported in part
by the National Natural Science Foundation of China (60806041,
60877055, and 60806002) and the Shanghai Leading Academic Discipline
Program (T0104). X. Chen is also supported in part by Shanghai
Educational Development Foundation (2007CG52) and Shanghai
Rising-Star Program (08QA14030).

\end{document}